\documentclass[prl,preprint,groupedaddress,showpacs]{revtex4}

\usepackage{amsmath}
\usepackage{amssymb}
\usepackage{amsfonts}
\usepackage{mathrsfs}
\usepackage{epsfig}
\usepackage{bm}

\begin{document}

\title{The ion motion in self-modulated plasma wakefield accelerators}

\author{J. Vieira$^1$}
\email{jorge.vieira@ist.utl.pt}
\author{R. A. Fonseca$^{1,2}$}
\author{W.B. Mori$^3$}
\author{L. O. Silva$^1$}
\email{luis.silva@ist.utl.pt}

\affiliation{$^1$GoLP/Instituto de Plasmas e Fus\~{a}o Nuclear-Laborat\'orio Associado,  Instituto Superior T\'{e}cnico, Lisboa, Portugal}
\affiliation{$^2$DCTI/ISCTE Lisbon University Institute, 1649-026 Lisbon, Portugal}
\affiliation{$^3$Department of Physics and Astronomy, UCLA, Los Angeles, California 90095, USA}


\today

\begin{abstract}
The effects of plasma ion motion in self-modulated plasma based accelerators is examined. An analytical model describing ion motion in the narrow beam limit is developed, and confirmed through multi-dimensional particle-in-cell simulations. It is shown that the ion motion can lead to the early saturation of the self-modulation instability, and to the suppression of the accelerating gradients. This can reduce the total energy that can be transformed into kinetic energy of accelerated particles. For the parameters of future proton-driven plasma accelerator experiments, the ion dynamics can have a strong impact. Possible methods to mitigate the effects of the ion motion in future experiments are demonstrated.  
\end{abstract}

\maketitle

Plasma based accelerators (PBA)~\cite{bib:tajima_prl_1979} sustain large amplitude waves that can trap and accelerate particles to high energies in distances more than three orders of magnitude shorter than conventional devices~\cite{bib:patel_nature_2007}. Currently, PBAs use $\lesssim 10~\mathrm{J}$ laser beams~\cite{bib:leemans_nat_2006} (LWFA), and $\lesssim 1~\mathrm{kJ}$ electron and positron bunches~\cite{bib:chen_prl_1985,bib:blumenfeld_nat_2007,bib:blue_prl_2003} (PWFA) to accelerate $\simeq~$1 GeV (LWFA) to $\simeq~$100 GeV (PWFA) electron bunches. The use of 100 kJ proton bunch drivers to generate TeV-class electrons was recently proposed in the so called proton-driven plasma-wakefield accelerator (PDPWFA)~\cite{bib:caldwell_natphys_2009} which relies on very short proton bunches ($\sigma_z < 10~\mathrm{\mu m}$), with more than $10^{11}$ protons per bunch to excite plasma waves in the blowout regime~\cite{bib:pukhov_apb_2002}.

The shortest proton bunches currently available, however, are long, with $\sigma_z \gtrsim 10~\mathrm{cm}$. Thus, a future PDPWFA experiment~\cite{bib:caldwell_arxiv_2011} will operate in weakly relativistic regimes where two stream-like (self-modulational) instabilities~\cite{bib:lawson_book} dominate the proton beam dynamics and wakefield excitation~\cite{bib:kumar_prl_2010,bib:vieira_pop_2012}. There are many analogies between self-modulation (S-M)of lasers and particle beams~\cite{bib:mori_ieee_1997}, and this scenario is also of relevance for the propagation of intense plasma streams in astrophysics~\cite{bib:silva_aip_2006}.

Recent work on the S-M of long particle bunches~\cite{bib:caldwell_arxiv_2011,bib:kumar_prl_2010} revealed that the wake amplitude and phase velocity can be nearly constant once the instability saturates. However, it was assumed that the background plasma ions are immobile. This will certainly not be the case for very long beams when $\omega_{pi} \sigma_z / c \gg 1$ ($\omega_{pi} = \sqrt{4 \pi n_0 e^2/m_i}$ is the ion plasma frequency, $e$ the elementary charge, and $m_i$ the ion mass). It is therefore important to understand the role of the ion motion in the S-M of very long beams, i.e. $\omega_{pi} \sigma_z/c = 1.88 \times 10^2 (\sigma_z/10~\mathrm{cm}) \sqrt{m_e/m_i} \sqrt{n_0 [\mathrm{cm}^{-3}]}$, where $m_e$ is the electron mass, and to determine conditions for which the plasma ion motion can be minimized.

Previous work addressed the role of ion motion due to the space charge forces of ultra-intense driving electron bunches in the PWFA nonlinear blowout regime~\cite{bib:rosenzweig_prl_2005}. Here we explore the motion of the plasma ions in the linear regime, where the ions respond to the plasma wave ponderomotive force.

In this Letter, we show that the background plasma ion motion can play a key role in the S-M of proton beams. An analytical model, confirmed by multi-dimensional particle-in-cell (PIC) simulations in OSIRIS~\cite{bib:fonseca_book}, demonstrates that ions move due to the nonlinear plasma wave ponderomotive force on plasma electrons. Future PDPWFA experiments will operate in the narrow beam limit for which there has been little work, and where fluid theory may fail due to phase mixing. We thus used the plasma sheet model~\cite{bib:dawson_pr_1959} to describe wake excitation in this limit and derive exact expressions for the ponderomotive force. Full-scale simulations in conditions of the PDPWFA reveal that the ion motion suppresses the transverse self-modulation instability (SMI), and reduces the accelerating wakefields throughout most of the proton bunch. Preventing the ion motion is therefore crucial, in particular for a future PDPWFA experiment, and heavy ion plasmas (e.g. Ar$^+$) are required to prevent it.

We start with the plasma fluid equations. The linearized continuity equation for the ions is $\partial n_{i1}/\partial t + n_{0}\nabla \cdot \mathbf{v}_{i1}=0$, where $n_{i}=n_{0}+n_{i1}$ is the ion density, $n_0$ the background plasma density, $n_{i1}\ll n_{0}$ and $\mathbf{v}_{i1}\ll c$ the perturbed ion density and velocity respectively, and $t$ the time. We assume cold ions ($T_i=0$) such that the linearized Euler's equation is $\mathrm{d} \mathbf{v}_{i1}/\mathrm{d} t = Z e \mathbf{E}/m_i$, where $\mathbf{E}$ is the electric field, and Z the ion charge. To obtain $\mathbf{E}$ we ignore the electron inertia and assume that $n_i = n_e$ which is true if $\lambda_d^2 \nabla^2 (n_e/n_0) \ll 1$, giving $e \mathbf{E} = k T_e \nabla (n_i / n_0) + \mathbf{F}_p$ where $\mathbf{F}_p$ is the nonlinear ponderomotive force on the electron, $T_e$ the electron temperature, $\lambda_d = \sqrt{k T_e/4 \pi n_0 e^2}$ the Debye length, $k$ the Boltzmann's constant, and $\nabla^2$ the Laplacian operator. Differentiating the ion continuity equation, substituting the $\mathrm{d} \mathbf{v}_{i1} / \mathrm{d} t$ from the Euler's equation, and using $\mathbf{E}$ from the electrons' Euler equation then gives:
\begin{equation}
\label{eq:iondensity}
m_i \left[c^2 \frac{\partial^2 }{ \partial \xi^2} -c_s^2 \nabla^2\right] n_{i1} = -n_0 Z \nabla \cdot \mathbf{F}_p,
\end{equation}
where $c_s = \sqrt{Z k T_e / m_i}$ is the ion sound speed, and where $(\xi=z-ct,\mathbf{r}=\mathbf{r},\tau=t)$. Equation~(\ref{eq:iondensity}) describes the evolution of the ion density in a warm plasma with $n_{i1}\ll n_0$. For a plasma wave with frequency $\omega_p$, then $\mathbf{F}_p = e^2/(4 m_e \omega_p^2)\nabla \mathbf{\hat{E}}^2$~\cite{bib:silva_pre_1999}, where $\mathbf{\hat{E}}$ is the envelope of the plasma wave electric field.

We consider cylindrically symmetric drivers such that $\mathbf{E}$, $n_{e}$, and $n_i$ only depend on the radius $r$, on $z$, and on $t$. In general both radial and longitudinal components of $\mathbf{F}_p$ should be included~\cite{bib:pukhov_pc_2011}. For cigar-shaped drivers with $\sigma_z \gg \sigma_r$, however, $|\mathbf{F}_{p\perp}|\gg |\mathbf{F}_{p\|}|$ since $\partial_z \simeq \frac{1}{\sigma_z} \ll \partial_r \simeq \frac{1}{\sigma_r}$. In addition, for narrow beams with $\sigma_r \ll c/\omega_p$, $E_z\ll E_r$ since the motion of the plasma electrons is preferentially along $r$ and not along $z$. Under these assumptions, and for a cold plasma, Eq.~(\ref{eq:iondensity}) reduces to:
\begin{equation}
\label{eq:ni_simplified}
m_i c^2 \frac{\partial^2 n_{i1}}{\partial \xi^2} = -\frac{n_0 Z e^2}{4 m_e \omega_p^2} \nabla_r^2 \hat{E}_r^2.
\end{equation}
Furthermore for narrow and small amplitude wakes the plasma response is dominated by the electrostatic forces for $E_r$, so we can obtain $\mathbf{F}_p$ for narrow bunches using Dawson's sheet model~\cite{bib:dawson_pr_1959} where non-relativistic, electrostatic radial oscillations of cylindrically symmetric electron rings are considered. In the absence of trajectory crossing the equation of motion for an electron ring is~\cite{bib:dawson_pr_1959}:
\begin{equation}
\label{eq:eom} 
\frac{\mathrm{d}^2 r}{\mathrm{d} \xi^2} = -\frac{\omega_p^2 r}{2} + \frac{\omega_p^2 r_0^2}{2 r} - \frac{e E_{r}^{\mathrm{b}}}{m_e},
\end{equation}
where $r$ is the radial position of the electron ring, and $r_0$ its initial radial position. The first term on the right hand side of Eq.~(\ref{eq:eom}) corresponds to the ion channel focusing field ($E_r^i$), the second to the electrostatic repulsion from the plasma electrons ($E_r^e$) %
inside the ring (the charge inside the ring of electrons remains fixed if the electron rings do not cross), %
and the third to the radial force associated with a cylindrically symmetric particle beam with electric field $E_{r}^{\mathrm{b}}$. We let $r=r_0 + \Delta r(r_0,\xi)$ and expand the right hand side of Eq.~(\ref{eq:eom}) into powers of $\Delta r/r_0$ yielding:
\begin{eqnarray}
\label{eq:eom_harmonic}
c^2\frac{\mathrm{d}^2 \Delta r}{\mathrm{d} \xi^2} & = & - \frac{e E_{r}^{\mathrm{b}}(r_0,\xi)}{m_e} - \\
& - &  \omega_p^2 \Delta r \left(1+\frac{e \nabla_{r_0} E_{r}^{\mathrm{b}}(r_0)}{\omega_p^2 m_e}\right) +\mathcal{O}\left(\Delta r\right)^2, \nonumber
\end{eqnarray}
where the derivatives are evaluated at $r=r_0$. Equation~(\ref{eq:eom_harmonic}) is also valid in 2D slab geometry. We consider a flat top driver in $\xi$ with length $\sigma_z$. Inside the beam ($0<\xi<\sigma_z$), $\Delta r = A_<\left(\cos \phi - 1\right)$ where  $A_{<}(r_0)=\frac{ E_{r}^{\mathrm{b}}(r_0)}{m_e \omega_p^2/e + \nabla_{r_0}E_{r}^{\mathrm{b}}(r_0)}$, and $\phi = (\omega_p \xi/c) \sqrt{1+(e/\omega_p^2 m_e) \nabla_{r0}E_{r}^{\mathrm{b}}(r_0)}$. Behind the driver ($\xi>\sigma_z$), $\Delta r = A_> \cos \phi$, where $A_{>}(r_0)=\left[\Delta r^{2} + c^2/\omega_p^2 (\mathrm{d} \Delta r/\mathrm{d} \xi)^2 \right]^{1/2}$, $\Delta r=r(\sigma_z)-r_0$, $\phi = \omega_p \xi/c + \phi_0$, and $\phi_0$ is the phase at $\xi=\sigma_z$. These trajectories are in agreement with PIC simulation results using narrow drivers ($\sigma_r \ll c/\omega_p$).

To determine $E_r(r)$ self-consistently the expression for the electron trajectories is inserted in $E_r\left(r_0(r,\xi),r\right)=E_r^i\left(r_0(r,\xi),r\right)+E_r^e\left(r_0(r,\xi),r\right)+E_r^{\mathrm{b}}(r)$ if $r_0(r,\xi)$ was known. To determine $r_0(r,\xi)$, $r=r_0+\Delta r (r_0,\xi)$  is inverted using the Lagrange's implicit function theorem. To lowest order we find $r_0(r,\xi)$ by Taylor expanding $r(r_0)$ in powers of $\Delta r$ yielding $r \simeq \left[r-A_<(\cos\phi-1)\right] - [1+\nabla_r A_<(\cos\phi-1)]\Delta r$ for $\xi<\sigma_z$, and $r \simeq r - A_>\cos\phi + (1+\nabla_r A_> \cos\phi ) \Delta r$ for $\xi>\sigma_z$, and then by solving these expressions for $r_0$. Inserting the result into the right hand side of Eq.~(\ref{eq:eom_harmonic}) yields:
\begin{eqnarray}
\label{eq:eradial_inside}
E_r^{\xi<\sigma_z} = \frac{\hat{E}_{r<} (-1+\cos\phi)}{1+\nabla_r {e \hat{E}_{r<}}/(m_e \omega_p^2)(\cos\phi-1)},
\end{eqnarray}
for $\xi<\sigma_z$ and :
\begin{eqnarray}
\label{eq:eradial_outside}
E_r^{\xi>\sigma_z} = \frac{\hat{E}_{r>} \cos\phi}{1+\nabla_r {e \hat{E}_{r>}}/(m_e \omega_p^2)\cos\phi},
\end{eqnarray}
for $\xi>\sigma_z$, where $\hat{E}_{r\gtrless}= m_e \omega_p^2 A_{\gtrless}/e$ is the amplitude of the radial plasma wave. When $\hat{E}_{r\gtrless} (e/m_e \omega_p) \ll 1$, and $\nabla_r \hat{E}_{r\gtrless} (e/m_e \omega_p) \ll \omega_p/c$, $E_r$ is purely sinusoidal. However, when narrow drivers are used, $\nabla_r \hat{E}_{r\gtrless} (e/m_e \omega_p) \gtrsim \omega_p/c$ and the wake becomes anharmonic even though $\hat{E}_{r\gtrless} (e/m_e \omega_p) \ll 1$. Excellent agreement was found between the analytical model and PIC simulations~\cite{bib:vieira_inpreparation} using narrow drivers, where linear fluid theory fails to accurately describe wake excitation.

The average $E_r$ over a plasma oscillation is finite. This can be interpreted as the electrostatic force from the ions that is required to balance a nonlinear average force pushing electrons outward. Thus, $F_{p\perp}=|\mathbf{F}_{p\perp}|$ is simply obtained by averaging $E_r$ over one oscillation, i.e. $e\langle E_r\rangle = F_{p\perp}$. Using the fact that $\int_{0}^{2\pi} \left[1+x (\cos\phi-1) \right]^{-1} \mathrm{d}\phi = -(1 + 2 x)^{-1/2}$, $\int_{0}^{2\pi} \cos\phi/\left[1 + x (\cos\phi-1) \right]^{-1} \mathrm{d}\phi = (1+x-\sqrt{1+2 x})/(x \sqrt{1+2 x})$, and $\int_{0}^{2\pi} \cos\phi / \left(1+x \cos\phi \right)^{-1} \mathrm{d}\phi = [1-(1-x^2)^{-1/2}]/x$~\cite{bib:integrals}, then the average value of Eq.~(\ref{eq:eradial_inside}) for $\xi<\sigma_z$, and near the axis becomes :
\begin{eqnarray}
\label{eq:eradial_ave_inside}
\langle E_r \rangle^{\xi<\sigma_z} = \frac{m_e \omega_p^2 r}{e} \left[1-\frac{1}{\sqrt{1-\frac{2 e}{m_e \omega_p^2} \nabla_r \hat{E}_{r<}}}\right]+ E_r^b,
\end{eqnarray}
and the average value of Eq.~(\ref{eq:eradial_outside}) for $\xi>\sigma_z$ at any $r$ becomes :
\begin{eqnarray}
\label{eq:eradial_ave_outside}
\langle E_r \rangle^{\xi>\sigma_z} = \frac{m_e \omega_p^2 }{e}\frac{\hat{E}_{r>}}{\nabla_r \hat{E}_{r>}}\left(1-\frac{1}{\sqrt{1 - \frac{e^2}{m_e^2 \omega_p^4} (\nabla_r \hat{E}_{r>})^2}}\right).
\end{eqnarray}

Equations~(\ref{eq:eradial_ave_inside},\ref{eq:eradial_ave_outside}) generalize the fluid theory nonlinear ponderomotive force ($F_p=e\langle E_r\rangle$) for narrow plasma waves and are correct to lowest order in $r-r_0$, i.e. they are valid as long as $\hat{E}_{r\gtrless}$ is calculated for sinusoidal trajectories. The corrections due to anharmonic oscillations in $\Delta r$ could also be included by considering higher order terms in Eq.~(\ref{eq:eom_harmonic}). To make a connection to the usual expression for $F_p$ we make a Taylor series expansion for small $\nabla_r \hat{E}_r$ in Eq.~(\ref{eq:eradial_ave_outside}) which yields $F_{p\perp}=- e^2/(4 m_e \omega_p^2) \nabla_r \hat{E}_r^2 + \mathcal{O}\left(\hat{E}_r^4\right)$. Equations~(\ref{eq:eradial_ave_inside}-\ref{eq:eradial_ave_outside}) differ from $F_{p\perp}^{\mathrm{fl}}$ when the wakefield becomes nonlinear; the wake can be anharmonic even when $e/(m_e \omega_p c)\hat{E}_r \ll 1$ if $e/(m_e \omega_p^2)\nabla_r \hat{E}_r \lesssim 1$, if $\sigma_r \ll c/\omega_p$~\cite{bib:vieira_inpreparation}. This is the limit of relevance for the PDWPFA.

The ion density perturbations in the wake of proton bunches can now be found by substituting $\langle E_r\rangle$ into Eq.~(\ref{eq:ni_simplified}). For $\xi\ll c/\omega_{pi}$ this yields $n_i = n_0\left[1+ \frac{e Z \xi^2}{2 m_i c^2} \nabla \cdot \langle E_r \rangle\right]$. Near the axis where $\hat{E}_r\simeq r \nabla_r \hat{E}_r|_0$, then, for $\xi<\sigma_z$, $n_i^{\xi<\sigma_z}=n_{i0} [1-\frac{m_e Z \omega_p^2 \xi^2}{m_i c^2} (1-(1-\frac{2 e}{m_e \omega_p^2} \nabla_r \hat{E}_{r<})^{-1/2}+\frac{e}{m_e \omega_p^2}\nabla_r E_r^b)]$. For $\xi>\sigma_z$, at any $r$, $n_i^{\xi>\sigma_z}= n_0\left[1 - \frac{m_e Z \omega_p^2 \xi^2}{m_i c^2} \left(1-(1-e^2(\nabla_r \hat{E}_{r>})^2/m_e^2 \omega_p^4)^{-1/2}\right)\right]$. Thus, $F_{p\perp}$ pulls the plasma ions towards the axis. Ion density voids are also formed near the edge of the wakefield.

\begin{figure}[htbp]
\begin{center}
\includegraphics[width=\columnwidth]{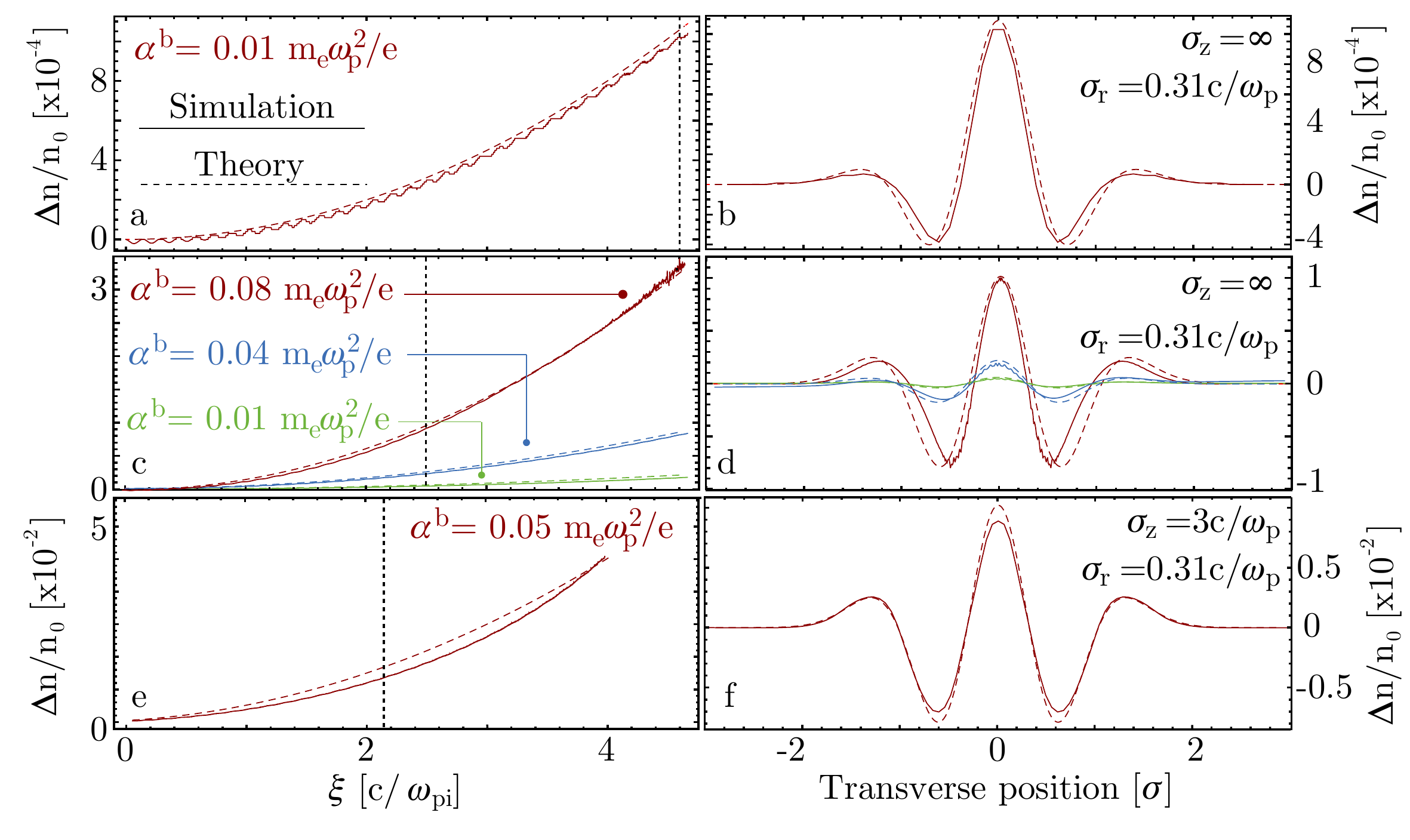}
\caption{\label{fig:comparison} Comparison between the ion density profile ($\Delta n = n_i-n_0$) analytical predictions (dashed lines) with 3D (a-b) and 2D (c-f) OSIRIS PIC simulations (solid lines). (a,c,e) 3D on-axis plasma proton density profile. (b,d,f) corresponding transverse density profiles at the $\xi$ of the vertical dashed lines.}  
\end{center}
\end{figure}

These theoretical predictions were compared with 2D slab geometry and 3D PIC simulations. The simulations used a 3D (2D) window moving at $c$, with dimensions $200 \times 4 \times 4 (c/\omega_p)^3$ ($400 \times4 (c/\omega_p)^2$), and a grid with $2560\times 512\times 512$ ($10240\times 1024$) cells in the longitudinal and transverse directions respectively. Each cell contains $2\times 1\times 1$ ($2 \times 2$) electrons and ions with mass $m_i=1836~m_e$. The simulations considered $E_r^{\mathrm{b}}= \alpha^{\mathrm{b}} r \exp\left[-r^2/\sigma_r^2\right]$. Both finite flat-top and infinite drivers were used. The values for $\alpha^{\mathrm{b}}$ and $\sigma_r$ are given in Fig.~\ref{fig:comparison}. Figure~\ref{fig:comparison} shows the simulations and the predictions from the analytical model (Eqs.~(\ref{eq:iondensity}), (\ref{eq:eradial_ave_inside},\ref{eq:eradial_ave_outside})) (where $\nabla^2=\mathrm{d}^2/\mathrm{d} r^2$ was used in 2D). There is excellent agreement for this range of parameters. Additional simulations with smaller $\sigma_r$ confirmed that fluid theory underestimates the ion density modulations whereas the predictions based on the ring model reproduce the simulation results accurately~\cite{bib:vieira_inpreparation}.

If the wake is excited by a self-modulated particle beam then it will grow in $\xi$ back through the bunch. If the S-M is constant then the wake will grow secularly in $\xi$. Under these conditions $E_{r\mathrm{wake}}\gg E_{r\mathrm{beam}}$ so the ponderomotive force can be obtained from Eq.~(\ref{eq:eradial_ave_outside}). We compared the predictions for $n_i$ using the full value for $\langle E_r\rangle$ in Eq.~(\ref{eq:eradial_ave_outside}) and the first term in the expansion $e\langle E_r\rangle \simeq -e^2/(4 m_e \omega_p^2)\nabla_r \hat{E}_r^2$ for $\sigma_z=100 c/\omega_p$, $n_b= 0.01-0.04 n_0$ and $\sigma_r\ll 1$. We assumed that $\hat{E}_r$ grew secularly with a resonant driving term of $E_{rb}$. We find that keeping only the first term in the expansion underestimates the ion fluctuations, $\delta \equiv (n_i-n_0)/n_0$ in a wide range by 2-45\%.

As a guide, one can estimate the position within the beam for which the ion compression on-axis is smaller than an accepted value for $\delta$. Assuming secular growth for $\hat{E}_r$, this occurs when $\xi_{\mathrm{crit}}/\sigma_z\lesssim 1$ where $\xi_{\mathrm{crit}}$
is the required distance for $\delta \simeq 1$ at $r=0$. In self-modulated regimes the onset of the ion motion then occurs when:
\begin{equation}
\label{eq:onset_general}
\frac{\xi_\mathrm{crit}}{\sigma_z} = \left(\frac{m_i c^2}{m_e Z \sigma_z^2 \omega_p^2}\right)^{1/2}\left(\frac{4 \pi m_e \omega_p^2}{e \nabla E_r^b} - \frac{3 e \sigma_z^2 \nabla E_r^b}{8 \pi m_e c^2}+\mathcal{O}(E_r^{b})^2\right)
\end{equation}
Equation~(\ref{eq:onset_general}) shows that the ion motion can be minimized in the presence of heavier plasma ions (i.e. lower $m_i/Z$). The leading order term coincides with the fluid theory result, and the remaining terms are corrections associated with the generalized ponderomotive force. For the expected initial parameters of the PDPWFA, where $n_0=10^{14}~\mathrm{cm}^{-3}$, $n_b=10^{12}~\mathrm{cm}^{-3}$, and $\sigma_z\simeq 12~\mathrm{cm}$, Eq.~(\ref{eq:onset_general}) shows that the ion motion is not important since $\xi^{\mathrm{ion}}/\sigma_z\gg 1$. However, for self-modulated beams $\xi_{\mathrm{crit}}/\sigma_z\gtrsim 1$, and the plasma ion motion can become important. To lowest order, the onset of ion motion beyond $\xi=\sigma_z$ can be reached by using more massive ions such that $[m_i/m_e]^{\mathrm{SM}} \gtrsim 0.08 \sigma_z^4[\mathrm{cm}] n_b^2 [10^{12}~\mathrm{cm}^{-3}]/\delta$. For a $\delta = 0.05$, a mass ratio of $[m_i/m_e]^{\mathrm{SM}} \gtrsim 40\times 10^3$ is required. This indicates that singly ionized Argon plasmas (or heavier) could be used to avoid the deleterious effects of the ion dynamics.

We have performed fully self-consistent simulations of the PDPWFA using 2D cylindrically symmetric geometry. The simulations used a moving window with dimensions $680 \times 8 (c/\omega_p)^2$, divided into $13600 \times 320$ cells, with $2\times2$ electrons and ions per cell. Plasmas with $m_i/m_e = 1836$ ($\mathrm{H}^+$), $m_i/m_e = 73440$ ($\mathrm{Ar}^+$), and $m_i = \infty$ (immobile ions) were considered with $n_0=10^{14}~\mathrm{cm}^{-3}$. An SPS-LHC CERN proton bunch  was initialized with an energy of 450 GeV and with a half-cut density profile given by $n_b = n_{b0} \left[1+\cos{\left(\sqrt{\pi/2} (z-z_0)/\sigma_z\right)}\right] \exp\left[-r^2/(2 \sigma_r^2)\right]$ for $0<z<z_0=\sigma_z\sqrt{2 \pi}$, with $n_{b0}/n_0=0.0152$, $\sigma_z=225.6~c/\omega_p=12~$cm, and $\sigma_r = 0.376~c/\omega_p=200~\mu$m.

\begin{figure}[htbp]
\begin{center}
\includegraphics[width=\columnwidth]{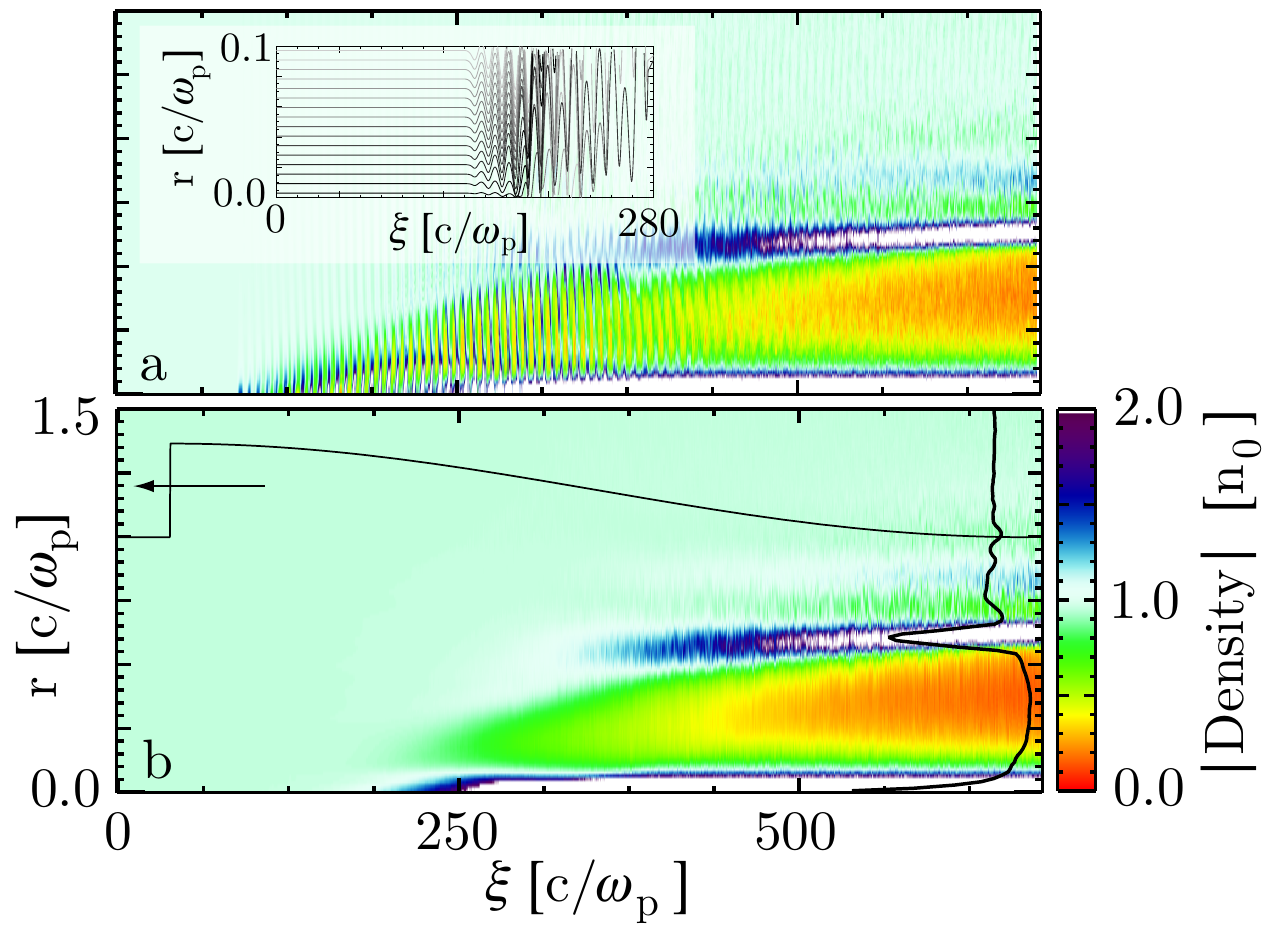}
\caption{\label{fig:hydrogen} 2D cylindrically symmetric OSIRIS simulations of a PDPWFA at $\tau=11450 / \omega_p$ (6.1 m) in an Hydrogen plasma. (a) Plasma electron density. The inset shows sample electron trajectories colored according to the initial radius (b) Corresponding plasma ion density. The horizontal solid line represents the driving ion beam density profile. The arrow indicates the propagation direction. The vertical solid line is a lineout of $n_i$ at $\xi=640 c/\omega_p$.}  
\end{center}
\end{figure}

\begin{figure}[htbp]
\begin{center}
\includegraphics[width=\columnwidth]{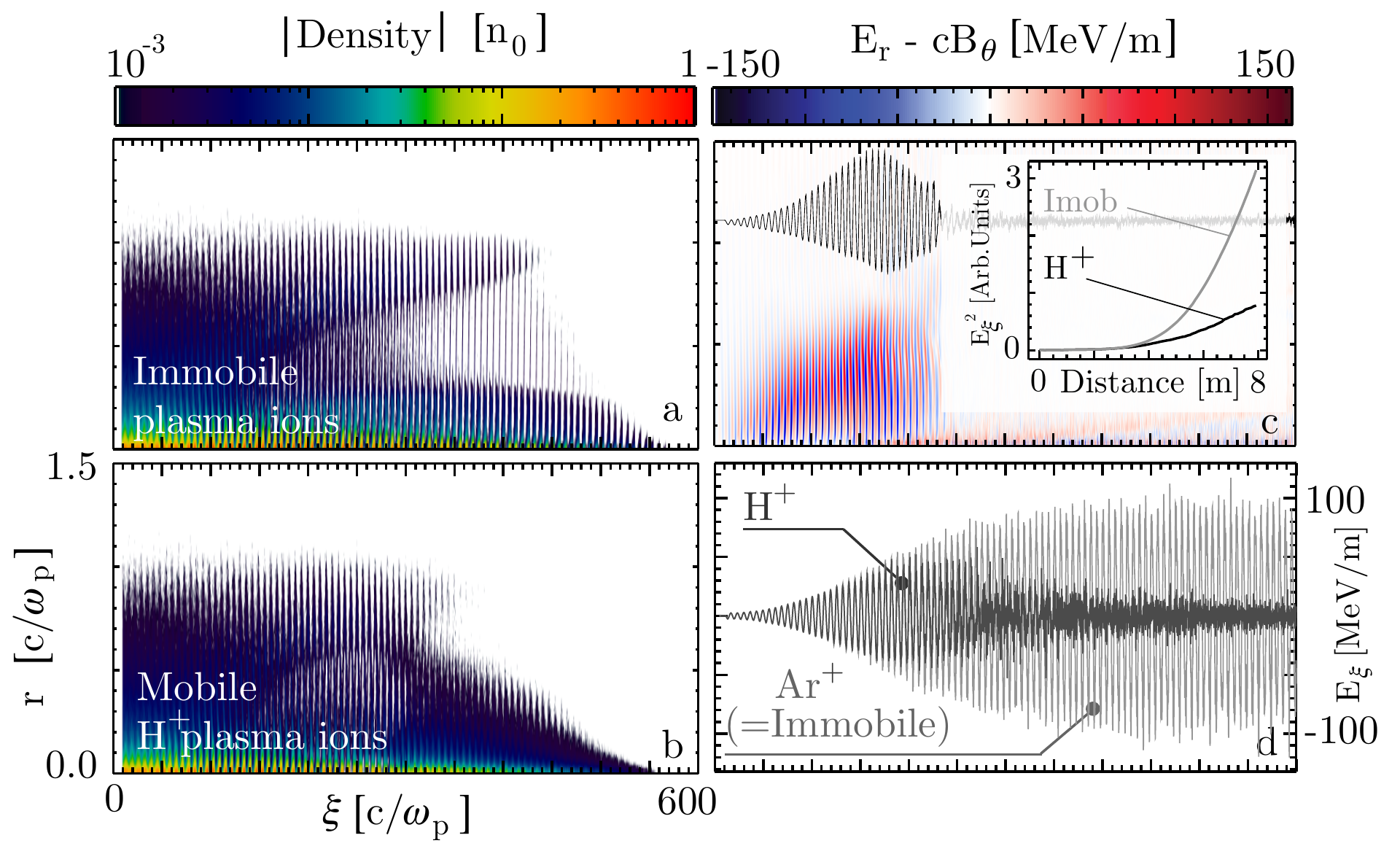}
\caption{\label{fig:self_modulation} OSIRIS simulation results of a PDPWFA at $\tau=11450 / \omega_p$ (6.1 m). (a-b) show the proton bunch density. (c) Plasma focusing force profile with mobile plasma ions. The solid line is a lineout at $r=0.5c/\omega_p$. The inset shows the evolution of the energy of $E_{\xi}$. (d) on-axis accelerating gradients using mobile $\mathrm{Ar}^+$ and $\mathrm{H}^+$ ions.}  
\end{center}
\end{figure}

The simulations confirm that the motion of the plasma ions can be neglected until the bunch significantly self-modulates. Once the bunch self-modulates, the (H plasma) ion compression strongly modifies the wakefields. The plasma electron and ion density after the beam has been self-modulated are shown in Figs.~\ref{fig:hydrogen}a-b. Ions have not moved significantly at the front of the proton bunch ($\xi\lesssim 250 c/\omega_p$). However, for $\xi\gtrsim 250 c/\omega_p$, the on-axis ion density grows such that $n_i/n_{0}\gg 1$. Ion density voids, where $n_i/n_{0}\simeq 0$, are also formed near $r\simeq \sigma_r$. The strong ion motion leads to electron trajectory crossing (Fig.~\ref{fig:hydrogen}a). The electron flow becomes turbulent, the plasma reaches a quasi-neutral state, and the wakefield disappears for $\xi\ge 400~c/\omega_p$ (Fig.~\ref{fig:self_modulation}).

If a H plasma is used the SMI slows down or becomes suppressed at the back of the beam ($\xi\gtrsim 250c/\omega_p$) because the plasma becomes quasi-neutral and fields become smaller. This is why the proton bunch density modulations are less pronounced in Fig.~\ref{fig:self_modulation}b than in Fig.~\ref{fig:self_modulation}a. Although the bunch is fully self-modulated in Fig.~\ref{fig:self_modulation}a some of the protons are still inside the box as revealed by the structures at $r\sim c/\omega_p$. The reason for the slowdown/suppression of the SMI is also seen in Fig.~\ref{fig:self_modulation}c showing that the SMI driving field $E_r-B_{\theta}$ vanishes at the back of the bunch. The amount of energy that can be transferred to accelerated particles is also smaller when ion motion occurs (inset of Fig.~\ref{fig:self_modulation}c). The inset of Fig.~\ref{fig:self_modulation}c also illustrates the slowdown of the SMI when plasma ion motion occurs. Similarly to the focusing force, and also because the plasma is quasi neutral, the accelerating fields drop abruptly for $\xi\gtrsim 250 c/\omega_p$, i.e. when plasma ions move. This further reduces the proton bunch energy modulations, and the energy that can be transferred to accelerated particles. Little ion motion occurs when $\mathrm{Ar}^+$ ions are used, and the SMI and $E_{\mathrm{accel}}$ develop as if ions are immobile. This is in excellent agreement with our theoretical results. These results strongly suggest that a future PDPWFA experiment could be performed with Ar plasmas. There are additional factors that may also change the growth of the instability including detuning due to strong energy chirps along the beam, or the presence of plasma density ramps \cite{bib:schroeder_pop_2012}. In realistic conditions, however, these effects are not as important as the plasma ion motion. 

In conclusion, we showed that the plasma ion dynamics can strongly affect a future PDPWFA experiment. It causes the early saturation of the self-modulation instability, reduces the accelerating gradients, and hence limits the energy transfer from the driver to accelerated particles. The conditions to minimize the impact of the ion motion were identified. This work demonstrates that the deleterious effects associated with the plasma ion motion can be avoided by resorting to plasmas with higher ion charge to mass ratios.

\begin{acknowledgements}
The authors acknowledge fruitful discussions with Prof. A. Pukhov. Work supported by the European Research Council (ERC-2010-AdG Grant 267841), by FCT (Portugal) through grants SFRH/BPD/71166/2010, PTDC/FIS/111720/2009, and CERN/FP/116388/2010; DE-FC02-07ER41500 and DE-FG02-92ER40727 US and NSF PHY--0904039 and PHY-0936266. Simulations were done on the IST Cluster at IST, Jaguar supercomputer under INCITE, on the Jugene supercomputer in Germany under PRACE.
\end{acknowledgements}

\end{document}